\documentclass[aps,pra,twocolumn,superscriptaddress,showpacs,nofootinbib]{revtex4}
\usepackage{epsfig}
\usepackage{amssymb}
\usepackage{bm}
\usepackage{epsf}
\DeclareFontFamily{OT1}{rsfs}{}
\DeclareFontShape{OT1}{rsfs}{m}{n}{<5> rsfs5 <7> rsfs7 <10> rsfs10}{}
\DeclareSymbolFont{mathrsfs}{OT1}{rsfs}{m}{n}
\DeclareSymbolFontAlphabet{\mathrsfs}{mathrsfs}

\newcommand{\A}{\mathbf{A}}

\newcommand{\K}{\mathbf{k}}
\renewcommand{\P}{\mathbf{p}}

\newcommand{\R}{\mathbf{r}}

\newcommand*{\ket}[1]{\left| #1 \right\rangle}
\newcommand*{\bra}[1]{\left\langle{#1}\right|}

\renewcommand{\Im}{{\rm Im}\,}

\begin{document}

\title{An electron in the presence of multiple zero range potentials and an external laser field -- exact solutions for photoionization and stimulated bremsstrahlung}

\author{Denys I. Bondar}
\email{dbondar@sciborg.uwaterloo.ca}
\affiliation{University of Waterloo, Waterloo, Ontario N2L 3G1, Canada}
\affiliation{National Research Council of Canada, Ottawa, Ontario K1A 0R6, Canada}

\author{Ryan Murray}
\email{ryan.murray@nrc.ca}
\affiliation{University of Waterloo, Waterloo, Ontario N2L 3G1, Canada}
\affiliation{Imperial College, London SW7 2BW, U.K.}

\author{Misha Yu. Ivanov}
\email{m.ivanov@imperial.ac.uk}
\affiliation{Imperial College, London SW7 2BW, U.K.}

\date{\today}

\begin{abstract}
The method of zero range potential (ZRP) for one-electron problems is reviewed. In the absence of an external electromagnetic field, the notion of a ZRP is introduced from different points of view and for an arbitrary dimension of space. Then, three-dimensional problems of motion of an electron in the field of several ZRPs and laser radiation are studied. Exact wave functions for the processes of photoionization and stimulated bremsstrahlung in the presence of the laser field of an arbitrary pulse shape are obtained in the form of one-dimensional integral representations.
\end{abstract}


\maketitle

\section{Introduction}

The zero range potential (ZRP) method is widely employed by physicists, mathematicians, and chemists \cite{Lieb1966, Drukarev1978, Demkov1988, Kleber1994a, Albeverio2005} because it usually allows one to obtain exact solutions\footnote{
Following  Ref. \cite{Kleber1994a}, we call a solution exact if it is expressed either in an analytical form or as an integral.
}
for a broad variety of model systems. The term ZRP may sound bizarre for beginners because a potential is usually represented by a non-vanishing  function in some region. An example of the ``function'' being nonzero at a single point $\R_0$ is the Dirac $\delta$-function, $\delta(\R-\R_0)$. Thus, the ZRP has various other names such as, e.g., ``point interaction,''  ``delta interaction,'' ``contact interaction.''  But the $\delta$-function is not the only example. In general,  an arbitrary ``function'' concentrated at $\R_0$ can be represented as $\sum_{n} c_n\delta^{(n)}(\R-\R_0)$ \cite{Gelfand1964}. However, all these objects are not ordinary functions; they are distributions and require special mathematical care \cite{Gelfand1964}. (We will return to this way of tackling a ZRP in Sec. \ref{Sec1s4}.)

The term ZRP should not be taken too literally. In fact, the ZRP method should be understood rather as a schematic or axiomatic description of the interaction of a particle with an object, which we call the ZRP. Note that there is no standardized definition of the ZRP; the definition varies for different problems. The essence of the ZRP method is to define what we mean exactly by the ZRP in a given situation. In the current paper, we pursue this point of view on the ZRP and demonstrate different approaches to define such a notion and interrelations among these definitions. Such a presentation should be beneficial especially for neophytes.

Kronig and Penney \cite{Kronig1931} were the first to employ ZRPs. They formulated the well known Kroning-Penney model of solids. A few years later, Bethe and Peierls \cite{Bethe1935} and Thomas \cite{Thomas1935} used a single ZRP in the theory of a deuteron. The first application of ZRPs in atomic physics was made by Fermi \cite{Fermi1934}. He also considered neutron scattering in substances containing hydrogen by means of ZRPs \cite{Fermi1936}. The first rigorous mathematical treatment of the notion of a ZRP was given by Faddeev and Berezin \cite{Berezin1961}. Regarding further development of the method of ZRP see, e.g., Refs. \cite{Drukarev1978, Demkov1988, Albeverio2005} and references therein.

The method of the ZRP entered strong field physics when analytical results on ionization of a weakly-bounded electron (negative ions), i.e., bounded in the field of a ZRP, were reported \cite{Nikishov1966, Perelomov1966a, Perelomov67a}. The exact three-dimensional solution to the problem of ionization of a particle in a ZRP by circularly polarized laser radiation was found in Refs. \cite{Manakov1976, Berson1975}. Afterwards, ionization in 
a elliptically polarized monochromatic laser field has been solved exactly in Refs. \cite{Manakov1979, Manakov1980}. Stimulated bremsstrahlung radiation in the presence of a ZRP was studied in Refs. \cite{Berson1975, Faisal1989a}. The exact solution for the one-dimensional Kronig-Penney model in a monochromatic laser field was achieved in Refs. \cite{Faisal1989}. High-harmonic generation from a model atom with a ZRP was investigated in Ref. \cite{Becker1990}. The ZRP method was used to model quasimolecular ions in a circularly polarized laser field in Ref. \cite{Krsti1991}. The exact solution for the problem of laser-assisted scattering from a one-dimensional $\delta$-function potential was reached in Refs. \cite{Faisal1989a, LaGattuta1994}. Decay of a weakly bound level in a monochromatic electromagnetic field and a static magnetic field was studied in Ref. \cite{Rylyuk2003a}. Photoionization of a negative ion in a bichromatic laser field was analyzed in Refs. \cite{Kiyan1989, Pazdzersky1997}. Recently, the ZRP method and its generalizations have been remarkably useful in understanding rescattering, above threshold ionization, and high harmonic generation \cite{Borca2002, Frolov2003a, Frolov2003, Flegel2005, Frolov2006, Frolov2008a, Frolov2009a, Frolov2009}. A more complete list of applications of ZRP in strong field physics  can be found in reviews \cite{Manakov2000, Manakov2003}.

Despite intense research in this direction, the problem of photoionization of  a weakly bounded electron by the laser field of an arbitrary vector potential, $\A(t)$, has not been solved exactly. Not only do we solve such a problem in the current paper, but we also obtain solutions for photoionization as well as stimulated bremsstrahlung of an electron bounded in the field of several ZRPs. The latter one is not only of an academic importance. Processes induced in molecular systems by an external laser field is a hot topic in strong field physics (for review see, e.g., Ref. \cite{Krausz2009}). Evidently, Coulomb forces ought to play an important role in these processes. Nonetheless, identification of features caused by the Coulomb interactions is a challenging task in general case. Our solution would allow to answer a complementary question: ``which processes are not due to the long range part of the Coulomb forces ?''  Indeed, the system of an electron moving in the field of many ZRPs is perhaps the simplest molecular model without long range forces.

It is noteworthy to mention that the ZRP method is not only fruitful for modeling a broad variety of so-called weakly bounded systems, but also it has been utilized to solve approximately the Schr\"{o}dinger equation for some class of potentials \cite{Subramanyan1970, Sigel1972}. Furthermore, the $\delta$-function as a potential term in the Schr\"{o}dinger equation has the following interesting property: the Hamiltonian in this equation has the minimum lowest eigenvalue among all potentials of a given ``area'' \cite{Keller1961}. Hence,  the convergence of variational calculations may be improved by including solutions of the corresponding ZRP problem into the variational basis set (see, e.g., Ref. \cite{Demkov1988}).

The structure of the current paper is as follows. As it was mentioned at the beginning of the Introduction, the notion of a ZRP is multifaceted; an attempt to manifest its richness by presenting different definitions of a ZRP is made in Sec. \ref{Sec1}. Presenting the matter of Sec. \ref{Sec1}, we often sacrifice mathematical rigor and technical details in favour of a clear and physically intuitive presentation; additionally, we present references to literature where more detailed discussions can be found. In general, Sec. \ref{Sec1} can be regarded as an overture to Sec. \ref{Sec2}, which is the main part of this paper. Problems of photoionization and stimulated bremsstrahlung of a one electron system in the field of ZRPs are discussed and solved exactly  in Secs. \ref{Sec2s1} and \ref{Sec2s2}, correspondingly.

Finally, for the sake of clarity, we mention a few words regarding notations and conventions used thorough the paper. Let $d$ denote a dimension of the space. Many equations are valid for an arbitrary dimension of the space; thus unless stated otherwise, the positive integer $d$ is not specified. Bold letters (e.g., $\R$, $\K$, $\P$) denote $d$-dimensional vectors; $r\equiv |\R|$ -- the absolute value of $d$-dimensional vector $\R$. $\delta(\R)$ is the $d$-dimensional Dirac $\delta$-function. $N$ denote the number of ZRPs, and unless stated otherwise its value is not fixed. Vectors ${\bf R}_j$ ($j=1,2,\ldots, N$) specify the locations of the ZRPs. Atomic units, $\hbar = m_e = |e| =1$, are used throughout.

\section{A ZRP without the presence of  a laser field}\label{Sec1}

\subsection{Physically intuitive method}\label{Sec1s1}

Probably most straightforward and intuitive way of introducing a ZRP is to replace the Schr\"{o}dinger equation inside the ZRP by a certain boundary condition on the wave function at the center of the ZRP.  This boundary condition can be postulating after analyzing solutions of the Schr\"{o}dinger equation for a ``shrinking'' potential. Let us present an example of such a description \cite{Baz1971, Demkov1988, Perelomov1998}.

Assuming the three-dimensional space, we consider a particle in an s-state in the presence of a spherically symmetric potential well defined as $r_j < b$, where $\R_j = \R - {\bf R}_j$ is the distance between the position of the particle $\R$ and the center of the well ${\bf R}_j$ and $b$ is the radius of the well. The wave function of the particle outside the potential well has the following form
\begin{eqnarray}\label{WaveFunSStateSpherWell3D}
\psi(\R) = c_j \exp(-\alpha_j r_j)/r_j, \qquad (r_j > b) 
\end{eqnarray}
where $c_j$ is the normalization constant and $\alpha_j = \sqrt{2I_p}$, $I_p$ is the binding energy (ionization potential). Now we contract the width of the well ($b\to 0$)  and simultaneously increase its depth  ($U_0 \propto 1/b^2$) such that $I_p$ as well as $\alpha_j$ remains constant. Hence, the potential well is replaced by a boundary condition at ${\bf R}_j$:
\begin{eqnarray}\label{BoundaryCondSingleZRP3D}
\psi(\R)\stackrel{r_j \to 0}{\longrightarrow} c_j\left[ 1/r_j - \alpha_j + O(r_j)\right].
\end{eqnarray}
Having done this limiting procedure, we can define the ZRP as follows: a wave function $\psi(\R)$ of a particle moving in the field of a ZRP located at ${\bf R}_j$ is a solution of the stationary Schr\"{o}dinger equation, $-\Delta\psi(\R)/2 = E\psi(\R)$, subjected to boundary  condition (\ref{BoundaryCondSingleZRP3D}). Note that parameter $\alpha_j$ does not depend on the energy of the particle $E$; $\alpha_j$ is to be regarded as merely a characteristic of the ZRP and $2\pi/\alpha_j$ is sometimes called the depth of the ZRP \cite{Demkov1988} or the renormalized coupling constant (see Sec. \ref{Sec1s2}).

One might have a methodological concern regarding boundary condition (\ref{BoundaryCondSingleZRP3D}): can a wave function have a singularity in quantum mechanics? Indeed, the postulates of quantum mechanics do not forbid for a wave function to be singular  as long as the probability of finding the particle in a volume $V$, 
\begin{eqnarray}\label{ProbInt}
{\rm Pr}\left( \R\in V\right) = \int_V |\psi(\R)|^2 d^3\R,
\end{eqnarray}
which is a measurable quantity, is continuous. Since the integral in Eq. (\ref{ProbInt}) exists if $V$ is a small ball centered at ${\bf R}_j$, wave functions satisfying boundary condition (\ref{BoundaryCondSingleZRP3D}) are allowed. Being more rigorous, we recall the first postulate of quantum mechanics: that wave functions are described by normalized vectors of a Hilbert space. The Hilbert space under discussion is $\mathrsfs{L}_2(\mathbb{R}^3)$ -- the set of functions $\phi(\R)$ such that $\int |\phi(\R)|^2d^3\R$ is finite; if $V=\mathbb{R}^3$, then Eq. (\ref{ProbInt}) is the definition of the norm of such a Hilbert space. In other words,  $\mathrsfs{L}_2(\mathbb{R}^3)$ is a set of bound states. Thus, it is noteworthy to recall the fact that  scattering states, which are solutions of the stationary Schr\"{o}dinger equation subjected to outgoing or ingoing wave boundary conditions at infinity, are not elements of the Hilbert space $\mathrsfs{L}_2(\mathbb{R}^3)$ simply because they are not square integrable. Mathematically speaking, the scattering states are distributional eigenfunctions of the Hamiltonian (recall that they are ``normalized'' to the $\delta$-function).

The presented limit transition can be performed in the one- and two-dimensional cases. The boundary condition for a wave function $\psi(x)$ in the one-dimensional case is given by
\begin{eqnarray}\label{BoundaryCondSingleZRP1D}
\left.\frac{d\psi}{dx}\right|_{x=X_j + 0} - \left.\frac{d\psi}{dx}\right|_{x=X_j - 0} = -2\alpha_j \psi(X_j),
\end{eqnarray}
where $X_j$ denotes the position of the ZRP; the boundary condition for the two-dimensional case reads 
\begin{eqnarray}\label{BoundaryCondSingleZRP2D}
\psi(\R)\stackrel{r_j \to 0}{\longrightarrow} c_j\left[ \ln r_j + \ln(\alpha_j/2) + C + O(r_j^2\ln r_j)\right],
\end{eqnarray}
where $C = 0.5772\ldots$ being the Euler-Mascheroni constant.

Further important generalizations are possible. Boundary conditions (\ref{BoundaryCondSingleZRP3D}), (\ref{BoundaryCondSingleZRP1D}), and (\ref{BoundaryCondSingleZRP2D}) remain unaltered even if we add a potential without a singularity at ${\bf R}_j$ into the stationary Schr\"{o}dinger equation. In the case of many ($N$)  ZRPs, we have $N$ boundary conditions for $j=1,2,\ldots, N$ (constants $c_j$ as well as $\alpha_j$ may differ for different $j$).

The stationary two-center problem within the ZRP method ($N=2$) has been pioneered by Smirnov and Firsov \cite{Smirnov1965}; afterwards, the general solution for the problem of a particle motion in a combine fields of several ZRPs has been found  \cite{Adamov1971, Dalidchik1974}. Moreover, there have been obtained a broad variety of Green's functions for such a problem (see, e.g., Refs. \cite{Maleev1966, Dalidchik1973, Labzovskii1973}). Regarding current status of the usage of the method of ZRP to model molecular systems see, e.g., Refs. \cite{Hogrevea2009}.

Different boundary conditions also have been obtained and applied to  more elaborate cases, e.g., such as a particle in the presence of an external static magnetic field \cite{Adamov1971, Rebane1976}, the time-dependant formulation of detachment of an electron from a negative ion in collision with an atom \cite{Demkov1964}, and many others. A more complete list of applications can be found in Refs \cite{Drukarev1978, Demkov1988}.

\subsection{Self-adjoint extensions of a Hermitian Hamiltonian}\label{Sec1s2}

This section is an effort to build the bridge between two stylistically different points of view on the same object -- mathematicians' and physicists' way of defying a ZRP.

According to the postulates of quantum mechanics, measurable quantities are represented by a {\it self-adjoint} operators, which act on a Hilbert space of wave functions of a given system. Note that there is a subtle difference between a Hermitian and self-adjoint operator. Usually, this difference is not important for physicists; however, it plays a crucial role in the theoretical formulation of a ZRP. 

Let us briefly clarify the difference. A linear operator $\hat{A}$, acting on a Hilbert space $\mathrsfs{H}$, is a linear mapping $\hat{A}: D(\hat{A})\to \mathrsfs{H}$, where $D(\hat{A})$ is a domain of the operator -- a set of elements on which the action of operator is defined. (In general, the domain $D(\hat{A})$ does not coincide with the entire Hilbert space $\mathrsfs{H}$.) The adjoint operator $\hat{A}^{\dagger}$ of an operator $\hat{A}$, whose domains are $D(\hat{A})$ and $D(\hat{A}^{\dagger})$ correspondingly, is such that the following equality is satisfied
\begin{eqnarray}
(\hat{A}^{\dagger}f, \, g ) = (f, \hat{A}g), \qquad \forall f,g \in D(\hat{A}).
\end{eqnarray}
The operator is self-adjoint if $\hat{A}^{\dagger} = \hat{A}$, i.e., the action of $\hat{A}^{\dagger}$ coincides with the action of $\hat{A}$, and $D(\hat{A})\subset D(\hat{A}^{\dagger})$. A self-adjoint operator $\hat{A}$ is a Hermitian operator if $D(\hat{A})= D(\hat{A}^{\dagger})$. Note that there is no distinction between self-adjoint and Hermitian operators in a finite dimensional Hilbert space $\mathrsfs{H}$ because operators are given by matrices and the operation of multiplication of a finite matrix by a vector is defined for all vectors from $\mathrsfs{H}$, viz.,  $D(\hat{A})\equiv D(\hat{A}^{\dagger})\equiv \mathrsfs{H}$ for an arbitrary matrix/operator $\hat{A}$.

Now we summarize two important results from the theory of self-adjoint operators that are widely employed in physics often without being referred to. The purpose of such a summary is to clarify the postulate of quantum mechanics, i.e., to accentuate the importance for the Hamiltonian of a quantum system as well as other observables  to be represented by self-adjoint operators and not by Hermitian ones.  The first result is {\it the Hilbert-Schmidt theorem} \cite{Hutson2005}: if $\hat{A}$ is a compact self-adjoint operator, then its eigenfunctions form an orthonormal basis for the Hilbert space. Second, {\it Stone's theorem} \cite{Reed1980, Araujo2008}:  if  $\hat{U}(t)$ is a strongly continuous unitary group (which describes the evolution of a quantum system), then there is a unique self-adjoint operator $\hat{H}$ (the Hamiltonian of the given quantum system) such that
\begin{eqnarray}
i\frac{\partial}{\partial t}\hat{U}(t) f = \hat{H}\hat{U}(t)f, \qquad \forall f\in D(\hat{H}),
\end{eqnarray}
this equation is formally expressed as $\hat{U}(t) = \exp(-i\hat{H}t)$. Physically speaking, if the time-independent Hamiltonian $\hat{H}$ is a self-adjoint operator, then Stone's theorem guarantees the unique solution of the time-dependent Schr\"{o}dinger equation such that the normalization of the wave function is constant in time. (Regarding the generalization of Stone's theorem to the case of a time-dependent Hamiltonian see, e.g., Sec. X.12 of Ref. \cite{Reed1975}.) 

An important fact is that a Hermitian operator can be transformed into a self-adjoint one by redefining the domain of the operator, e.g., by means of imposing appropriate boundary conditions on the wave functions on which the operator acts. This procedure is an object of study of the theory of self-adjoint extensions of a Hermitian operator. We shall not discus this theory in the current paper. There are tutorials oriented for physicists that not only explain this theory, but also illustrate it on many physically interesting examples \cite{Capri1977, Zhu1993, Bonneau2001, Araujo2004, Araujo2008}. Regarding the review of the theory of extensions and exactly solvable models see, e.g., Ref. \cite{Pavlov1987}. Rigorous discussions of the theory can be found in mathematical textbooks \cite{Akhiezer1993, Naimark1967}. (There are also textbooks oriented for applications of functional analysis \cite{Hutson2005, Reed1975}.)

We describe the application of the theory of self-adjoint extensions of a Hermitian Hamiltonian to the problem of ZRP in a nutshell (for further details and more rigorous formulations see Ref. \cite{Albeverio2005} as well as tutorial \cite{Araujo2004}). This treatment of a ZRP by means of the theory of self-adjoint extensions was pioneered by Faddeev and Berezin \cite{Berezin1961}. Assuming that a single ZRP is located at the origin, we denote the Hamiltonian for a particle in the field of the ZRP by $\hat{H}_{ZRP}$ and $d=1,2,3$ -- the dimension of the space. A priori fact is that the Hamiltonian $\hat{H}_{ZRP}$ must coincide with the free particle Hamiltonian, $\hat{H}_0 = -\Delta/2$, in the space without the origin, $\mathbb{R}^d\setminus {\bf 0}$. Therefore, let us restrict the domain of $\hat{H}_0$ to $\mathrsfs{C}_0^{\infty}(\mathbb{R}^d\setminus{\bf 0})$ -- a set of infinitely differentiable functions that vanish outside a compact set which does not contain the origin (e.g., let $d=1$ then $\mathrsfs{C}_0^{\infty}(\mathbb{R}\setminus{\bf 0})$ consists of infinitely differentiable functions $f_{abcd}(x)$ that vanish for $x\in (-\infty, -a)\cup(-b, c)\cup(d, +\infty)$, where $a$, $b$, $c$, and $d$ are arbitrary positive numbers). One can verify that $\hat{H}_0$ defined on such a domain is a Hermitian but not self-adjoint operator. Applying the theory of extensions, one concludes that for $d=2,3$ there is a one-parameter family of self-adjoint extensions of the Hermitian operator $\hat{H}_0$ indexed by a ``renormalized coupling constant'' [and defined on an appropriate subset of $\mathrsfs{L}_2(\mathbb{R}^d)$.] These self-adjoint operators are to be postulated as definitions of $\hat{H}_{ZRP}$ in the two- and three-dimensional spaces. However, the one-dimensional case is special because the Hermitian operator $\hat{H}_0$ has a four-parameter family of self-adjoint extensions, which physically means that there are not only $\delta$-interactions [this alternative name for a ZRP comes from the fact that heuristically speaking, a one-dimensional ZRP can be represent as the term $-\mu\delta(x)$ in the Hamiltonian, $\hat{H}_{ZRP}=-\Delta/2 - \mu\delta(x)$], but also  new types of point interactions such as $\delta'$-interaction [which has a heuristic form $-\mu\delta'(x)$] and powers of the $\delta$-interaction \cite{Rosinger1978, Rosinger1980}. Note that these new interactions cannot be introduced by means of the physically intuitive method discussed above.

To draw the connection between this section and section \ref{Sec1s1}, we mention that the operator $\hat{H}_0$ defined on the set of functions that satisfy boundary conditions (\ref{BoundaryCondSingleZRP3D}), (\ref{BoundaryCondSingleZRP1D}), and (\ref{BoundaryCondSingleZRP2D}) is indeed self-adjoint. Unfortunately, book \cite{Demkov1988} as well as other physicists' publications demonstrates a lack of understanding of the last fact -- $\hat{H}_0$ is demonstrated to be a Hermitian operator (see page 6 of Ref. \cite{Demkov1988}), but no comment is made regarding its self-adjointness. This is not remarkable since physicists rarely distinguish between Hermitian and self-adjoint operators. Needless to say,  such a distinction is vital as far as the notion of a ZRP is concerned.

Now we clarify why Hamiltonians that contain the $\delta$-function as a potential term [e.g., $\hat{H}_{ZRP}=-\Delta/2 - \mu\delta(x)$] are not legitimate, and thus they are to be regarded as heuristic expressions.  The quantity $\mu\delta(\R)$, where $\mu\in\mathbb{R}$, is not an operator acting on the Hilbert space $\mathrsfs{L}_2(\mathbb{R}^d)$ since for an arbitrary $f(\R)\in\mathrsfs{L}_2(\mathbb{R}^d)$, the integral $\int |\delta(\R)f(\R)|^2 d^d \R$ does not make any sense. From this point of view, we want to emphasize that the approach of self-adjoint extensions of the free particle Hamiltonian does not assign any potential to the interaction of a particle with the ZRP, viz., the method of self-adjoint extensions gives us the ``potential-free'' interpretation of a ZRP.

\subsection{Renormalization Method}\label{Sec1s3}

A variety of renormalization techniques, some being similar to renormalization in quantum field theory, have been used in the problem of ZRPs \cite{Zeldovich1960, Berezin1961, Wodkiewicz1991, Jackiw1991, Krajewska2008} (see also Ch. X.11 of Ref. \cite{Albeverio2005}). In this section, we advocate a renormalization method, which combines advantages of known renormalization techniques. This method will be adapted to the time-dependent case in Sec. \ref{Sec2}.

We introduce our renormalization method on the example of $N$ ZRPs that are located at $\{ {\bf R}_j \}_{j=1}^N$, and the dimension of the space $d$ is not specified. We start from the time-independent Schr\"{o}dinger equation
\begin{eqnarray}\label{AbstrStacSchEq}
\left( \hat{\P}^2/2 + \hat{V} \right)\ket{\psi} = E\ket{\psi},
\end{eqnarray}
where $\hat{V}$ is the heuristic potential, which represents the ZRPs, and whose coordinate representation reads
\begin{eqnarray}\label{DeffV}
\bra{\R_1}\hat{V}\ket{\R_2} = -\delta(\R_1-\R_2)\sum_{j=1}^N\mu_j\delta(\R_1 - {\bf R}_j).
\end{eqnarray}
Projecting Eq. (\ref{AbstrStacSchEq}) onto the conjugate of a momentum eigenstate $\bra{\K}$, and using the equality $\bra{\K}{\bf R}_j \rangle = \exp(-i\K\cdot{\bf R}_j)/(2\pi)^{d/2}$, we obtain the Schr\"{o}dinger equation in the momentum representation
\begin{eqnarray}
\left( E - \K^2/2\right)\bra{\K}\psi\rangle = -\sum_{j=1}^N \beta_j \exp(-i\K\cdot{\bf R}_j ), \label{StatSchEqInMomentRepres}
\end{eqnarray}
where 
\begin{eqnarray}\label{DefBetaStac}
\beta_j = \mu_j \bra{{\bf R}_j}\psi\rangle /(2\pi)^{d/2}
\end{eqnarray} 
being a constant. Solution of Eq. (\ref{StatSchEqInMomentRepres}) depends on the sign of the energy $E$: if $E<0$, then there is a bound state solution
\begin{eqnarray}\label{StatBoundSolutionMomentRep}
\bra{\K}\psi\rangle = \sum_{j=1}^N \frac{\beta_j \exp(-i\K\cdot{\bf R}_j)}{\K^2/2 + |E|},
\end{eqnarray}
which will be an element of $\mathrsfs{L}_2(\mathbb{R}^d)$ after Fourier transformation. If $E$ is positive, then taking into account that $(\P^2 - \K^2)\delta(\P-\K)=0$, we obtain scattering state solutions
\begin{eqnarray}\label{StatScateringSolutionMomentRep}
\bra{\K}\psi\rangle = \delta(\P-\K) + \sum_{j=1}^N \frac{\beta_j \exp(-i\K\cdot{\bf R}_j)}{\K^2/2 -E \pm i0},
\end{eqnarray}
where $\P^2/2 = E$ and $\pm i0$ correspond to outgoing  and ingoing wave boundary conditions, respectively. Equation (\ref{StatScateringSolutionMomentRep}) clearly illustrates the fact mentioned in Sec. \ref{Sec1s1} that scattering state solutions are not elements of $\mathrsfs{L}_2(\mathbb{R}^d)$ -- they are distributions. 

Now we scrutinize bound state solution (\ref{StatBoundSolutionMomentRep}). Flourier transforming wave function (\ref{StatBoundSolutionMomentRep}), we obtain the wave function, $\psi(\R) \equiv \bra{\R}\psi\rangle$, in the coordinate representation in the following compact form (for an arbitrary spacial dimension $d$)
\begin{eqnarray}\label{StatBoundSolutionCoordRep}
\psi(\R) = \sum_{j=1}^N B_j \frac{K_{(d-2)/2}(\kappa r_j)}{r_j^{(d-2)/2}},
\end{eqnarray}
where $B_j$ are constants proportional to $\beta_j$, $\kappa = \sqrt{2|E|}\equiv\sqrt{2I_p}$, and $K_{\nu}(x)$ is the modified Bessel function of the second kind (or the Macdonald function \cite{Bateman1953}). We point out that the ZRP in higher dimensions ($d > 3$) is also of interest (see, e.g., Refs. \cite{Wodkiewicz1991, Grossmann1984}).
 Let us itemize interesting special cases of Eq. (\ref{StatBoundSolutionCoordRep})
\begin{eqnarray}
\psi(\R) = \left\{
\begin{array}{ccc}
\displaystyle \sum_{j=1}^N B_j \exp(-\kappa |x_j|), &\mbox{if}& d=1, \\
\displaystyle \sum_{j=1}^N B_j K_0(\kappa r_j), &\mbox{if}& d=2, \\
\displaystyle \sum_{j=1}^N B_j \exp(-\kappa r_j)/r_j, &\mbox{if}& d=3. 
\end{array}\right.
\end{eqnarray} 
In the case of a single ZRP ($N=1$), one can readily verify that wave function (\ref{StatBoundSolutionCoordRep}) indeed satisfies boundary conditions (\ref{BoundaryCondSingleZRP3D}), (\ref{BoundaryCondSingleZRP1D}), and (\ref{BoundaryCondSingleZRP2D}) for $d=1,2$, and $3$, correspondingly. In the case of many ZRPs, following Refs. \cite{Adamov1971, Dalidchik1974} (see also Ch. 3 of Ref. \cite{Demkov1988}), we impose the $N$ boundary conditions on wave function (\ref{StatBoundSolutionCoordRep}) at the points $\{\R = {\bf R}_j\}_{j=1}^N$ and obtain a homogeneous system of $N$ algebraic equations, from which the coefficients $\{B_j\}_{j=1}^N$ and $\kappa$ can be obtained. 

Now we shall return to the definition of the parameters $\beta_j$ [Eq. (\ref{DefBetaStac})]. Since $\bra{{\bf R}_j}\psi \rangle\equiv\psi({\bf R}_j)$ and $\psi({\bf R}_j)$ are singular for $d \geqslant 2$, we conclude that the parameters $\beta_j$ contain divergences in the case of $d \geqslant 2$.  
However, we have not used the freedom of choosing values of the coupling constants $\mu_j$. Hence, we interpret the values of  $\psi({\bf R}_j)$ as some infinite numbers, and thus set $\mu_j$ to be equal to corresponding infinitesimals such that $\beta_j\in\mathbb{R}$. Summarizing, we conclude that the normalization constant of the wave function ($B_j\propto \beta_j$) is renormalized within such a method. Note that the one-dimensional case is special because renormalization is not needed. 

The renormalization technique is generalized to describe a bound state with nonzero angular momentum in Appendix \ref{Appendix2}. 

\subsection{Miscellaneous methods}\label{Sec1s4}

In this section, we list other methods used to define the notion of a ZRP. Nevertheless, this list will not be exhaustive, and we refer the reader to monograph \cite{Albeverio2005} for other approaches. 

In Sec. \ref{Sec1s3}, we constructed the wave function of a particle in the field of ZRPs by employing heuristic potential (\ref{DeffV}) with infinitesimally small coupling constant $\mu$. However, one may still hope that it is possible to represent consistently a ZRP by means of the $\delta$-function with a non-vanishing coupling constant and additionally avoid the problem that the $\delta$-function does not define an operator on the Hilbert space (see Sec. \ref{Sec1s2}). As we have seen (e.g., Sec. \ref{Sec1s1}) that the wave function or/and its derivatives has a singularity at the point where a ZRP is located. Since the classical calculus is not applicable to the wave function at the singular point, we shall treat it as a distribution to have the properly defined operation of differentiation. If we follow such a  path we immediately run into another problem because the multiplication of distributions is not uniquely defined, and the stationary Schr\"{o}dinger equation contains the product of the wave function and the potential, which are distributions.  To illustrate the troublesomeness of multiplication of distributions, we recall the Schwartz counterexample,
$$
0 = 0\cdot 1/x = (\delta(x)\cdot x)\cdot 1/x \, \neq \, \delta(x)\cdot(x\cdot1/x) = \delta(x),
$$
i.e., there is no associative and commutative operation of multiplication consistent with multiplication by multipliers. The problem of multiplications of distributions is an active area of research in modern mathematical physics; some approaches to this problem can be found in Refs. \cite{Rosinger1978, Rosinger1980, Egorov1990, Colombeau1992, Danilov1998}. Summarizing, we say that for the construction of quantum mechanics with singular potentials, it might not be sufficient to regard functions corresponding to singular potentials as distributions due to onerousness of constructing a sufficiently large associative algebra of distributions.

Nevertheless, there have been successful attempts to construct axiomatically suitable algebras -- Shirokov's algebras \cite{Shirokov1979c, Shirokov1979b, Tolokonnikov1981, Tolokonnikov1982, Tolokonnikov1982a}, and  they have been used to  tackle the ZRP \cite{Shirokov1980, Talalov1981, Tsirova1981}. 
Antonevich  \cite{Antonevich1999} has also employed Colombeau's algebra of new generalized functions \cite{Colombeau1992} to the problem of ZRP. As a historical note, we mention that in the fifties,   Bogolubov suggested that a necessary attribute of any future local theory would be a modification of the subtractional formalism associated with the multiplication of distributions. Furthermore, there is a strong connection between regularization in quantum field theory and the problem of multiplication of distributions \cite{Bogoliubov1957, Bogoliubov1980}.

In Ref. \cite{DrozVincent2008}, a scalar product of some distributions  was introduced and used to define the one-dimensional ZRP as a singular potential bypassing the problem of multiplications of distributions. In some sense, this work is in spirit of Shirokov's algebras.

The renormalization technique (Sec. \ref{Sec1s3}) may also suggest that an elegant and satisfactory picture of ZRP could be achieved if the notions of ``infinite and infinitesimal quantities'' were formalized. This point of view leads to the nonstandard analysis, which can be looked upon as a rigorous mathematical realization of Newton's and Leibniz's ideas of infinitesimal quantities without running into contradictions. Non-standard analysis was introduced in the early sixties by Robinson \cite{Robinson1996}. (Non-mathematicians may find a comprehensible introduction to the nonstandard analysis in book \cite{Robert1988}.) In Refs. \cite{Albeverio1979, Albeverio1986}, the ZRP was quite organically defined within nonstandard analysis. Moreover, there have been attempts to formulate quantum mechanics in a non-standard Hilbert space \cite{Farrukh1975, Almeida2004}.

Fermi \cite{Fermi1934} indirectly proposed to replace the expression $\hat{V} = -\mu\delta(\R)$ in the three-dimensional case by
\begin{eqnarray}\label{FermiPotential}
\hat{V}_F = \frac{2\pi}{\kappa} \delta(\R)\frac{\partial}{\partial r}r \equiv \frac{2\pi}{\kappa} \delta(\R)\nabla\cdot\R,
\end{eqnarray}
where as previously $\kappa = \sqrt{2I_p}$, and we assumed that there is a single ZRP located at the origin. New heuristic potential (\ref{FermiPotential}) has been called the Fermi pseudopotential, and it has been widely exploited in physics (see, e.g., Refs. \cite{Wu1984, Faisal1989a, Wodkiewicz1991} and references therein). A main advantage of the Fermi pseudopotential compare to heuristic potential (\ref{DeffV}) is the fact that the coupling constant, $2\pi/\kappa$, is a real non-vanishing number. Nonetheless, the pseudopotential is still not an operator on the Hilbert space, similarly to $\hat{V}$ (see Sec. \ref{Sec1s2}).  Additionally, it elegantly avoids the problem of multiplication of distributions: since $\hat{V}_F$  acts on  a wave function with the asymptotic behavior (\ref{BoundaryCondSingleZRP3D}) near the origin, we obtain 
$$
\hat{V}_F \left[1/r + \kappa + O(r)\right] = 2\pi \delta(\R).
$$
However, the situation is opposite  once phenomena in an external electromagnetic field are considered: Let
$
\hat{H} = -\nabla^2 /2 + \hat{V}_F 
$
denote the Hamiltonian (in the coordinate representation) of a particle moving in the field of a ZRP modeled by the pseudopotential (\ref{FermiPotential}). As it was pointed out in Ref. \cite{Arrighini1986}, using the minimal coupling procedure, $-i\nabla\to -i\nabla + \A$, to ``turn on'' the electromagnetic field defined by the vector potential $\A = \A(\R, t)$, we update the form of the Hamiltonian $\hat{H}$ to
\begin{eqnarray}\label{FermiPseudoPotInField}
\hat{H} = \frac 12 \left[ -i\nabla + \A \right]^2 + \frac{2\pi i}{\kappa}\delta(\R)\left[-i\nabla + \A \right]\cdot\R.
\end{eqnarray}
If we assume that the Hamiltonian (\ref{FermiPseudoPotInField}) acts on wave functions that obey boundary condition  (\ref{BoundaryCondSingleZRP3D}) (or any singular boundary condition at the origin), then the last term of  Eq. (\ref{FermiPseudoPotInField}) indeed represents the multiplication of two distributions -- the $\delta$-function and the wave function. Note that the standard symbolic relationship, $\R\delta(\R)={\bf 0}$, is valid only when the functional $\R\delta(\R)$ acts on an infinitely differentiable function. (The latter fact nevertheless was not recognized in Ref. \cite{Arrighini1986}; besides, the existence of the last term of  Eq. (\ref{FermiPseudoPotInField}) was not even mentioned in Refs. \cite{Manakov1976, Berson1975, Manakov1979, Manakov1980, Faisal1989a, Manakov2000, Manakov2003}.) Therefore, we do not use the Fermi pseudopotential in the current paper; in fact, we firmly discourage one from utilizing it because, as it was seen in the previous illustration,  the Fermi pseudopotential is ``too'' {\it ad-hoc} to be  useful for the problems under scrutiny  in Sec. \ref{Sec2}.

\section{ZRPs in the presence of a laser field}\label{Sec2}

\subsection{Photoionization}\label{Sec2s1}

Similarly to scattering states (see Sec. \ref{Sec1s1}),  a wave function that describes photoionization is not normalizable, i.e., it is not an element of the Hilbert space $\mathrsfs{L}_2(\mathbb{R}^d)$. Hence, we cannot use the theory of self-adjoint extensions (Sec. \ref{Sec1s2}) to obtain boundary conditions, which should be satisfied by the wave function. Furthermore, the way of defying the ZRP as a limit of some continuous potentials is cumbersome analytically; nevertheless, one could attempt to use numerical solutions of the Schr\"{o}dinger equation with a time-dependent Hamiltonian to find such a definition. However, there is an easier way: A straightforward generalization of the renormalization technique presented in Sec. \ref{Sec1s3} allows us to define the problem of photoionization of a weakly bounded system (bounded in the field of ZRPs). 

From now onwards, we   solve the time-dependent Sch\"{o}dinger equation
\begin{eqnarray}\label{TimeDependSchEq}
i\frac{\partial}{\partial t}\ket{\Psi(t)} = \left[ \frac{\hat{\P}^2}2 + \hat{V} +\hat{\P}\cdot\A(t) + \frac 12 \A^2(t) \right]\ket{\Psi(t)},
\end{eqnarray}
where $\hat{V}$ is the heuristic potential, which represents the ZRPs, given by Eq. (\ref{DeffV}); $\A(t)$ is an arbitrary vector potential. A crucial step to solve such a problem is to make the following substitution into the  Sch\"{o}dinger equation 
\begin{eqnarray}\label{FloquetAnsatze}
\ket{\Psi(t)} = \exp({-i\epsilon t})\ket{\Phi(t)}.
\end{eqnarray}
Let the substitution (\ref{FloquetAnsatze}) be called the {\it Floquet ans\"{a}tze}; the reason of such a name will be discussed later in this section. The Sch\"{o}dinger equation for the wave function $\ket{\Phi}$ reads
\begin{eqnarray}
&& i\frac{\partial}{\partial t}\ket{\Phi(t)} = \hat{H}_{\epsilon}(t)\ket{\Phi(t)}, \label{SchrodEqForPhi}\\
&& \hat{H}_{\epsilon}(t) =\hat{\P}^2/2 + \hat{V} +\hat{\P}\cdot\A(t) +  \A^2(t)/2  - \epsilon, \nonumber
\end{eqnarray}
Partitioning the total Hamiltonian
\begin{eqnarray}
\hat{H}_{\epsilon}(t) = \hat{H}_V(t) + \hat{V},
\end{eqnarray}
we can write down the Lippmann-Schwinger equation in the post form  for the propagator $\hat{U}_{\epsilon}$ (see, e.g., Refs. \cite{Kleber1994a, Smirnova2007a})
\begin{eqnarray}\label{LippSchwEq}
\hat{U}_{\epsilon} (t, t_0) = \hat{U}_V(t,t_0) -i\int_{t_0}^t \hat{U}_V(t, t')\hat{V}\hat{U}_{\epsilon}(t', t_0)dt',
\end{eqnarray}
where the propagators $\hat{U}_{\epsilon}$ and $\hat{U}_V$ correspond to the Hamiltonian $\hat{H}_{\epsilon}$ and $\hat{H}_V$ respectively, viz.,
\begin{eqnarray}
\hat{U}_{\epsilon, \, V}(t, t') = \hat{T}\exp\left[ -i\int_{t'}^t \hat{H}_{\epsilon, \, V}(\tau)d\tau\right] ,
\end{eqnarray}
where $\hat{T}$ being the time ordering operator. The exact expression of the propagator $\hat{U}_V$ is known
\begin{eqnarray}
\hat{U}_V(t,t') = \exp[{i\epsilon(t-t')}] \hat{U}_{VG}(t,t'),
\end{eqnarray}
where $\hat{U}_{VG}(t,t')$ being the velocity gauge Volkov propagator 
\begin{eqnarray}\label{VolkovPropagator}
&& K_{VG}(\R, t |\R', t') \equiv \langle \R|\hat{U}_{VG}(t,t') |\R' \rangle = \nonumber\\
&& \qquad \frac {\theta(t-t')}{\left[2\pi (t-t')\right]^{d/2}}
\exp\Bigg\{ \frac i{2(t-t')}\Bigg[ \R-\R' \nonumber\\
&&\qquad -\int_{t'}^t \A(\tau)d\tau \Bigg]^2 - \frac i2 \int_{t'}^t \A^2(\tau)d\tau - i\frac {\pi d}4 \Bigg\}. \qquad
\end{eqnarray}
Since the problem of ionization is being considered, we introduce the following initial condition: the wave function $\ket{\Phi(t=-\infty)}=\ket{\psi}$ is a bound state [see, e.g., Eq. (\ref{StatBoundSolutionCoordRep})]. Substituting the following equalities 
\begin{eqnarray}
\ket{\Phi(t)} = \hat{U}_{\epsilon}(t, -\infty)\ket{\psi}, \quad
\lim_{t_0\to-\infty} \bra{\R}\hat{U}_V(t, t_0)\ket{\psi} = 0,  \nonumber
\end{eqnarray}
into Eq. (\ref{LippSchwEq}), we obtain the equation for the wave function, $\Phi(\R, t) \equiv \langle \R \ket{\Phi(t)}$,
\begin{eqnarray}
 \Phi(\R, t) &=& \sum_{j=1}^N G_j(\R), \label{IntRepOfPhi} \\
 G_j(\R) &=&  \int_{-\infty}^t  \frac{e^{i\epsilon(t-t')}}{(2\pi)^{-d/2}} K_{VG}(\R, t|{\bf R}_j, t') f_j(t')dt', \label{DeffGFunc}
\end{eqnarray}
where $f_j(t)$ is a time-dependent analogue of $\beta_j$ [Eq. (\ref{DefBetaStac})]
\begin{eqnarray}
f_j(t) = i\mu_j \Phi({\bf R}_j, t)/(2\pi)^{d/2}.
\end{eqnarray}

In the rest of the current section, we focus on the three-dimensional case. Cases of other dimensions can be tackled similarly. The asymptotic expansion of $G_j(\R)$ for small $r_j$ is found in Appendix \ref{Appendix1} [Eq. (\ref{AsymptGSmall3D})]. Now we are in position to postulate boundary conditions in the case of the presence of a laser field. We shall be guided by the two following principles: First, the boundary conditions ought to make Eq. (\ref{TimeDependSchEq}) solvable; second, the time-dependent boundary conditions must coincide with stationary boundary conditions (\ref{BoundaryCondSingleZRP3D}) once the laser field is off ($\A(t)\equiv {\bf 0}$). The following boundary conditions indeed fulfill these constrains 
\begin{eqnarray}\label{BoundaryCondPhotoionZRP3D}
\Psi(\R, t)\stackrel{r_j \to 0}{\longrightarrow} c_j(t)\left[ 1/r_j - \alpha_j +O\left(r_j\right) \right], \, j=1,\ldots,N,
\end{eqnarray}
where $c_j(t)$ are unknown functions of time, and $\alpha_j$ are characteristics of ZRPs, as in the field-free case. Therefore, we have reduce the problem of obtaining the wave function $\Psi$ to the problem of calculating the functions $c_j(t)$. These functions can be found once we impose the wave function given by Eq. (\ref{IntRepOfPhi}) to the boundary conditions (\ref{BoundaryCondPhotoionZRP3D}). Indeed, calculating the limits $r_j\to 0$ in Eq. (\ref{IntRepOfPhi}) and using Eqs. (\ref{BoundaryCondPhotoionZRP3D}) and (\ref{AsymptGSmall3D}), we achieve the system of integral equations for $f_j(t)$ ($j=1,\ldots,N$):
\begin{eqnarray}\label{SystemIntEqsFuncF3D}
&& \sqrt{2\pi}f_j(t)\left(i\alpha_j - \sqrt{2\epsilon}\right) = \sum_{k=1\atop k\neq j}^N G_k({\bf R}_j) \nonumber\\
&& + \int_0^{\infty} \frac{dx\, e^{i\epsilon x}}{(ix)^{3/2}}\left[ W(t, x)f_j(t-x) - f_j(t)\right],
\end{eqnarray}
where $W(t,x)$ is given by Eq. (\ref{WFuncDeff}). The functions $c_j(t)$ and $f_j(t)$ are connected by the equality: $c_j(t) = -i\sqrt{2\pi}\exp(-i\epsilon t)f_j(t)$. 

We summarize results obtained until now in the form of a definition: Photoionization of a one electron system bounded in the field of $N$ ZRPs, which are characterized by $\alpha_j$, is defined by the wave function $\Psi(\R,t)$
\begin{eqnarray}\label{TotalWaveFuncPhotoioniz}
\Psi(\R, t) = \exp(-i\epsilon t)\sum_{j=1}^N G_j(\R),
\end{eqnarray}
where $G_j(\R)$ is given by Eq. (\ref{DeffGFunc}), and the functions $f_j(t)$ and the parameter $\epsilon$ are determined by solving non-linear eigenfunction-eigenvalue problem (\ref{SystemIntEqsFuncF3D}).

However, solving non-linear eigenfunction-eigenvalue problems  is generally a challenging task (for review see, e.g.,  Ref. \cite{Golub2000}). In the case of monochromatic radiation, there exists the Floquet twin-transformation (see, e.g., Refs. \cite{Faisal1987, Faisal1989a}), which tremendously simplifies numerical solution of problem (\ref{SystemIntEqsFuncF3D}).

Note that all the consideration has been done only in the velocity gauge. Therefore, it is vital to check whether such a treatment is gauge invariant. It is well known that the wave function $\Psi(\R, t)$ is transformed as
\begin{eqnarray}\label{WaveFuncgaugeTrans}
\Psi(\R, t) \to \tilde{\Psi}(\R, t) = \exp[i\chi(\R, t)]\Psi(\R, t)
\end{eqnarray}
under some gauge transformation. Substituting Eq. (\ref{BoundaryCondPhotoionZRP3D}) into Eq. (\ref{WaveFuncgaugeTrans}), we obtain the boundary conditions for $\tilde{\Psi}(\R,t)$:
\begin{eqnarray}
\tilde{\Psi}(\R,t) \stackrel{r_j \to 0}{\longrightarrow} \tilde{c}_j(t)\left[ \frac 1{r_j} - \alpha_j +i\frac{\R_j\cdot\nabla\chi({\bf R}_j, t)}{r_j} + O\left(r_j\right) \right],
\end{eqnarray}
where $\tilde{c}_j(t) = \exp[i\chi({\bf R}_j, t)]c_j(t)$. Having applied our renormalization procedure to this new wave function, we obtain 
\begin{eqnarray}
\tilde{\Psi}(\R, t) = \exp[ -i\epsilon t + i\chi(\R, t)]\sum_{j=1}^N \tilde{G}_j(\R),
\end{eqnarray}
where $\tilde{G}_j(t)$ are given by Eq. (\ref{DeffGFunc}) after performing the substitution $f_j(t)\to\tilde{f}_j(t) = \exp[-i\chi({\bf R}_j, t)]f_j(t)$; the functions $\tilde{f}_j(t)$ satisfy Eq. (\ref{SystemIntEqsFuncF3D}) where $f_j(t)\to \tilde{f}_j(t)$ and $G_j(t)\to \tilde{G}_j(t)$; as in the velocity gauge, $\tilde{c}_j(t) = -i\sqrt{2\pi}\exp(-i\epsilon t)\tilde{f}_j(t)$. Roughly speaking, the gauge transformation factor is ``absorbed'' by  $\tilde{f}_j(t)$ such that all the equations are left form invariant. Hence, we have explicitly demonstrated that the developed technique is indeed gauge independent; as a result, it does not ``suffer'' from the ``curse of the displaced atom,'' which is an artifact associated with the choice of gauge in the strong field approximation \cite{Smirnova2007a}. 

We present the special case of  Eqs. (\ref{SystemIntEqsFuncF3D}) and (\ref{TotalWaveFuncPhotoioniz}) for a single ZRP ($N=1$) located at the origin (${\bf R}_1={\bf 0}$): The integral representation of the wave function $\Psi_{1}(\R, t)$ reads
\begin{eqnarray}
\Psi_{1}(\R, t) = (2\pi)^{3/2}\int_{-\infty}^t e^{-i\epsilon t'} K_{VG}(\R, t|{\bf 0}, t') f(t')dt',
\end{eqnarray}
where the unknown function, $f(t)$, and parameter, $\epsilon$, obey the integral equation
\begin{eqnarray}
&& -\sqrt{2\pi i}\left( \kappa + i\sqrt{2\epsilon}\right)f(t)  \nonumber\\
&& = \int_0^{\infty} \frac{dx\, e^{i\epsilon x}}{x^{3/2}}\left[ W(t, x)f(t-x) - f(t)\right].
\end{eqnarray}
This solution resembles the previous results \cite{Manakov1979, Manakov1980} in the case of a monochromatic laser field.

A vital step in our derivation was the Floquet ans\"{a}tze (\ref{FloquetAnsatze}). Note that in the case of ionization by a periodic laser field, this ans\"{a}tze is the result of the Floquet theorem, where $\epsilon$ is called the quasi energy and its imaginary part is proportional to the total ionization rates, and $\Phi(\R, t)$ being a time periodic function. In the case of a periodic field, the Floquet theorem as well as quasi energy formalism is extensively used in strong field physics, for reviews see, e.g., Refs.
\cite{Chu2004, Faisal1989a, Manakov1986, Rapoport1978}. Therefore, the Floquet ans\"{a}tze can be regarded as a trick to extend methods developed for periodic laser fields to the general case of arbitrary laser pulses.

\subsection{Stimulated Bremsstrahlung}\label{Sec2s2}

Once we have worked out photoionization, the problem of stimulated bremsstrahlung is trouble-free. In some sense, bremsstrahlung is even easer because there is no need to employ the Floquet ans\"{a}tze [Eq. (\ref{FloquetAnsatze})].

We want to find a solution of Eq. (\ref{TimeDependSchEq}) with the initial condition being an outgoing-wave   wave function (\ref{StatScateringSolutionMomentRep}), $\ket{\psi_{b}}$. Let $\ket{\Psi_b(t)}$ denote such a solution, and $\hat{U}$ represents the total propagator of the system at hand, $\ket{\Psi_b(t)}=\hat{U}(t,-\infty)\ket{\psi_b}$.  From the initial condition, we conclude that
\begin{eqnarray}\label{InitConditionBrem}
&& \lim_{t_0\to-\infty} \hat{U}_{VG}(t, t_0)\ket{\psi_b}  \nonumber\\
&& = (2\pi)^{d/2}\exp\left\{ -\frac i2\int^t\left[ \P +\A(\tau)\right]^2d\tau\right\}\ket{\P}. 
\end{eqnarray}
Substituting Eq. (\ref{InitConditionBrem}) into the Lippmann-Schwinger equation:
$$
\hat{U}(t,t_0) = \hat{U}_{VG}(t, t_0) -i\int_{t_0}^t \hat{U}_{VG}(t,t')\hat{V}\hat{U}(t', t_0)dt',
$$ 
the following integral representation of the total wave function is obtained
\begin{eqnarray}
&& \Psi_b(\R,t) = \exp\left\{ -\frac i2\int^t\left[ \P +\A(\tau)\right]^2d\tau +i\P\cdot\R\right\} \nonumber\\
&&\quad + (2\pi)^{d/2} \sum_{j=1}^N \int_{-\infty}^t K_{VG}(\R, t|{\bf R}_j, t')q_j(t')dt',
\end{eqnarray}
where $q_j(t) = i\mu_j \Psi_b({\bf R}_j, t)/(2\pi)^{d/2}$. In the three-dimensional case ($d=3$), the renormalized functions $q_j(t)$ obey the following system of non-homogeneous integral equations
\begin{eqnarray}
&& i\sqrt{2\pi}\alpha_j q_j(t) = \exp\left\{ -\frac i2\int^t\left[ \P +\A(\tau)\right]^2d\tau +i\P\cdot{\bf R}_j\right\} \nonumber\\
&&\qquad + (2\pi)^{3/2}\sum_{k=1 \atop k \neq j}^N \int_{-\infty}^t K_{VG}({\bf R}_j, t |{\bf R}_k, t')q_k(t')dt' \nonumber\\
&&\qquad + \int_0^{\infty} \frac{dx}{(ix)^{3/2}}\left[ W(t,x)q_j(t-x) - q_j(t)\right]. 
\end{eqnarray}

\appendix

\section{The renormalization technique for bound states of nonzero angular momentum}\label{Appendix2} 

In this appendix we restrict ourself to the case of there-dimensions ($d=3$) and a single ZRP centered at the origin (${\bf R}_j = {\bf 0}$). Then, the heuristic potential (\ref{DeffV}), $\bra{\R_1}\hat{V}\ket{\R_2} = -\mu\delta(\R_1-\R_2)\delta(\R_1)$, can be rewritten in the following form
\begin{eqnarray}\label{MomCoordDeffV}
\bra{\K}\hat{V}\ket{\R} = -\mu \sqrt{4\pi} Y_{00}(\Omega_{\K})\delta(\R),
\end{eqnarray}
where $Y_{lm}(\Omega_{\K})$ is the spherical harmonic, $\Omega_{\K}$ denote the pair of the spherical angular coordinates of the vector $\K$. Generalizing Eq. (\ref{MomCoordDeffV}), we define a new heuristic potential
\begin{eqnarray}
\bra{\K}\hat{V}_{lm}\ket{\R} = -\mu \sqrt{4\pi} k^l Y_{lm}(\Omega_{\K})\delta(\R).
\end{eqnarray}
Now, we solve the stationary Schr\"{o}dinger equation with this potential,
\begin{eqnarray}\nonumber
\left( \hat{\P}^2/2 + \hat{V}_{lm}\right)\ket{\psi} = E\ket{\psi},
\end{eqnarray}
The bound state solution reads
\begin{eqnarray}
\langle \K \ket{\psi} = -\frac{\bra{\K}\hat{V}_{lm}\ket{\psi}}{\K^2/2 + |E|}= \beta\frac{k^l Y_{lm}(\Omega_{\K})}{k^2 + \kappa^2},
\end{eqnarray}
where $\beta = 2\sqrt{4\pi}\mu\psi({\bf 0})$ is similar to Eq. (\ref{DefBetaStac}). Fourier transforming the wave function in the momentum representation, e.g., by employing the partial wave expansion, 
$$
\exp(i\K\cdot\R) = 4\pi\sum_{l',\, m'} i^{l'}j_{l'}(kr)Y_{l' m'}^*(\Omega_{\K})Y_{l' m'}(\Omega_{\R}),
$$
we obtain 
\begin{eqnarray}\label{WaveFunLStateSpherWell3D}
\psi(\R) = B \frac{K_{l+1/2}(\kappa r)}{\sqrt{r}} Y_{lm}(\Omega_{\R}),
\end{eqnarray}
where $B$ is a  constant proportional to $\beta$. The wave function (\ref{WaveFunLStateSpherWell3D}) also represents the wave function of a particle with an arbitrary angular momentum $l$ outside the spherical potential well. Evidently, Eq. (\ref{WaveFunLStateSpherWell3D}) coincides with Eq. (\ref{WaveFunSStateSpherWell3D}) for an s-state, $l=0$.

\section{The asymptotic behavior of the function $G_j(\R)$  for small $r_j$}\label{Appendix1}

In this appendix we calculate the asymptotics of the function $G_j(\R)$ [Eq. (\ref{DeffGFunc})]
when $\R\to {\bf R}_j$, i.e., $r_j\to 0$. Since an arbitrary propagator $K\left(\R, t | \R', t'\right)$ obeys the following normalization condition
\begin{eqnarray}\nonumber
\lim_{t\to t'} K\left(\R, t | \R', t'\right) = \delta\left( \R -\R'\right),
\end{eqnarray}
the function $G_j(\R)$ for any propagator $K$ may have a singularity at $r_j\to 0$. 

Let us separate this singularity. Having changed the variable of integration in Eq. (\ref{DeffGFunc}) to $x=t-t'$, we can represent $G_j(\R)$ as
\begin{eqnarray}\label{AppGNewVar}
&& G_j(\R) = f_j(t)e^{-i\pi d/4} \int_0^{\infty}\frac{dx}{x^{d/2}}\exp\left( \frac{i\R_j^2}{2x} + i \epsilon x \right)    \nonumber\\
&& + e^{-i\pi d/4}\int_0^{\infty}\frac{dx\, e^{i\epsilon x }}{ x^{d/2}}\Bigg[
\exp\Bigg\{\frac{i}{2x}\left( \R_j-\int_{t-x}^t\A(\tau)d\tau\right)^2 \nonumber\\
&& - \frac i2\int_{t-x}^t\A^2(\tau)d\tau \Bigg\}f_j(t-x) - f_j(t)\exp\left(\frac{i\R_j^2}{2x}\right) \Bigg]. 
\end{eqnarray}
One can readily notice that the expression under the second integral in Eq. (\ref{AppGNewVar}) is regular at $x=0$. Using the following integral representation of the Macdonald function \cite{Bateman1953}
\begin{eqnarray}\nonumber
K_{\nu}(\alpha z) = e^{i\nu\pi/2}\frac{\alpha^{\nu}}2 \int_0^{\infty} \exp\left[ \left(x-\frac{\alpha^2}{x}\right)\frac{iz}2 \right] x^{-\nu-1}dx,
\end{eqnarray}
such that $\Im(z)>0$ and  $\Im(\alpha^2z)>0$, we obtain  
\begin{eqnarray}\label{IntLaserFreeCase}
&&\int_0^{\infty}\frac{dx}{x^{d/2}}\exp\left( \frac{i\R_j^2}{2x} + i \epsilon x \right)  \nonumber\\
&&= 2\left( \frac{\sqrt{2\epsilon}}{r_j} \right)^{\frac{d-2}2} K_{(d-2)/2}\left(e^{-i\pi/2}\sqrt{2\epsilon}r_j\right), \quad
\end{eqnarray}
where the following convention has been adopted 
\begin{eqnarray}
z^c = |z|^c \exp(ic \arg z), \quad -\pi <\arg z \leqslant \pi.
\end{eqnarray} 
Substituting Eq. (\ref{IntLaserFreeCase}) into Eq. (\ref{AppGNewVar}), we find the sought asymptotic in the three-dimensional case 
\begin{eqnarray}\label{AsymptGSmall3D}
G_j(\R) &=& -i\sqrt{2\pi}f_j(t)\left( 1/r_j + i\sqrt{2\epsilon}\right) \nonumber\\
&& +  \int_0^{\infty} \frac{dx\, e^{i\epsilon x}}{(ix)^{3/2}}\left[ W(t, x)f_j(t-x) - f_j(t)\right] \nonumber\\
&& + O\left(r_j\right), \qquad (d=3)
\end{eqnarray}
where the function $W(t,x)$ is given by
\begin{eqnarray}\label{WFuncDeff}
W(t,x) &=& \exp\left\{\frac{i}{2x}\left(\int_{t-x}^t\A(\tau)d\tau\right)^2 \right. \nonumber\\
&&\qquad\qquad \left. 
- \frac i2\int_{t-x}^t\A^2(\tau)d\tau \right\},
\end{eqnarray}

\bibliography{ZRP}
\end{document}